\documentclass[prl,twocolumn,superscriptaddress]{revtex4-1}
\usepackage{graphicx}
\usepackage{amsmath,amssymb}
\usepackage[T1]{fontenc}
\usepackage{mathptmx}
\usepackage[dvipsnames]{color}
\definecolor{myblue}{named}{MidnightBlue}
\usepackage[a4paper=true,colorlinks=true,citecolor=myblue,linkcolor=black,urlcolor=myblue]{hyperref}

\newcommand{\ket}[1]{|#1\rangle}
\newcommand{\bra}[1]{\langle#1|}
\newcommand{\be}{\begin{equation}}
\newcommand{\ee}{\end{equation}}
\newcommand{\ba}{\begin{eqnarray}}
\newcommand{\ea}{\end{eqnarray}}
\usepackage{placeins}

\usepackage{txfonts}

\def\unity{\mathbb I}

\begin{document}

\title{Single-photon heat conduction in electrical circuits}

\author{P. J. Jones}
\affiliation{Department of Applied Physics/COMP, Aalto University, PO Box 14100, 00076 Aalto, Finland.}
\author{J. A. M. Huhtam\"aki}
\affiliation{Department of Applied Physics/COMP, Aalto University, PO Box 14100, 00076 Aalto, Finland.}
\author{K. Y. Tan}
\affiliation{Department of Applied Physics/COMP, Aalto University, PO Box 14100, 00076 Aalto, Finland.}
\author{M. M\"{o}tt\"{o}nen}
\affiliation{Department of Applied Physics/COMP, Aalto University, PO Box 14100, 00076 Aalto, Finland.}
\affiliation{Low Temperature Laboratory, Aalto University, PO Box
13500, 00076 Aalto, Finland.}

\begin{abstract}
We study photonic heat conduction between two resistors coupled weakly to a single superconducting microwave cavity.
At low enough temperature, the dominating part of the heat exchanged between the resistors is transmitted by single-photon excitations of the fundamental mode of the cavity. This manifestation of single-photon heat conduction should be experimentally observable with the current state of the art.
Our scheme can possibly be utilized in remote interference-free temperature control of electric components and environment engineering for superconducting qubits coupled to cavities.
\end{abstract}
\maketitle

In nanoscale devices where the structure size is comparable to the characteristic wavelength of the heat and charge carriers, quantum mechanical effects can have a great impact on their transport properties. An important manifestation of this phenomenon is the quantized electrical conductance~\cite{vanWees88, Wharam88}, i.e., the maximum current that an ideal conducting channel can support is limited by the conductance quantum. Also the thermal conductance is quantized, a result which can be derived from very general principles concerning the limit of information transfer~\cite{Rego98,Pendry83}. This general result implies that regardless of the nature of the carriers of thermal energy, there will be an upper bound on the amount of energy which can be transported by a single mode. This has been demonstrated experimentally for phonons~\cite{Schwab00} and electrons \cite{Chiatti06}. Recent experiments explicitly show that, as predicted in Ref.~\cite{Cleland04}, the quantization of thermal conductance applies equally to photons at low temperatures~\cite{Meschke06,Timofeev09}.
Further proposals have been presented for harnessing this photonic channel in useful applications~\cite{Ojanen08} and a convenient circuit model to compute the transmitted semiclassical power has been introduced~\cite{Pascal11}.

We extend these previous studies to the case of quantized heat transport mediated by single photons with a fixed frequency. This is achieved by using a superconducting cavity as the mediator for the heat. We derive the quantum mechanical interaction Hamiltonian between the cavity and the reservoirs, which allows us to describe the full quantum dynamics of the cavity modes under an engineered and tunable environment. The quantum model is found to agree extremely well with the semiclassical model for weak coupling and low temperatures. We note that the narrow bandwidth of the single-photon channel allows for interference-free remote temperature control of electronic components working at different frequency bands.


Microwave photons are extensively utilized in the various  implementations of superconducting quantum bits, qubits~\cite{Blais04,Majer07,Vion09,Simmonds07}. In these schemes microwave photons are employed, for example, as carriers for transporting information, or in order to perform control operations on the qubits. Hence, microwave cavities form an integral component of these superconducting computer architectures and already impressive progress has been made on generating, guiding and manipulating individual microwave photons within electrical circuits~\cite{GenPhotons,Leek10,Schoelkopf10,Cleland08}. We demonstrate how a resistor embedded in such a cavity may be used as a controllable environment for the superconducting qubits. The control allowed by our technique may therefore offer an effective method to gain further understanding and control~\cite{Valenzuela06, Ithier05} of the decoherence processes of cavity qubits~\cite{Houck08,Bertet05}.

\begin{figure}[htbp!]
\includegraphics[width=0.35\textwidth]{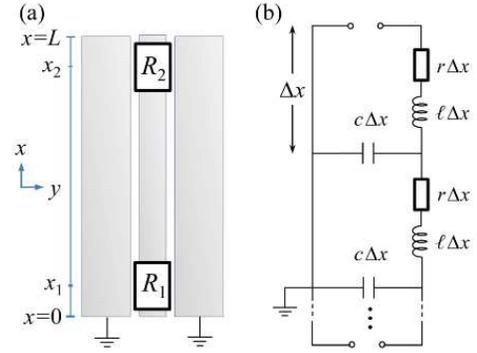} 

\caption{(color online) (a) Schematic figure of two resistors, $ R_{1}$ and $R_{2}$, in a coplanar waveguide cavity.
(b) The bare cavity can be modelled as a sequence of resistors and inductors with a capacitive connection to the ground. The values of $r$, $\ell$, and $c$ are the resistance, inductance, and capacitance per unit length, respectively. This discrete model approaches the true continuous case as $ \Delta x \rightarrow 0 $.}
\label{fig:CPW}
\end{figure}

The physical system we consider is shown schematically in Fig.~\ref{fig:CPW}(a).
 It consists of a superconducting coplanar waveguide microwave resonator \cite{Wallraff08} with two resistors embedded close to the opposite ends of the cavity. We first consider the case of a bare cavity without resistors. Such a transmission line may be represented as a sequence of resistors, inductors, and capacitors as shown in Fig.~\ref{fig:CPW}(b). We neglect first the intrinsic dissipation of the cavity. The quantum mechanical voltage operator for this one-dimensional representation of a transmission line can be found by quantizing the corresponding Euler--Lagrange equations in the continuum limit \cite{Blais04,Wallraff04}. With the ends of the cavity defined at $ x=0 $ and $ x=L $, and the boundary conditions imposed by charge neutrality, the resulting voltage operator may be written as $ \hat{V}_{\textrm{cav}} = \sum_{k=1}^{\infty} \hat{A}_{k}(t)\cos \left({k\pi x}/{L}\right) $, where $ \hat{A}_{k}(t)  = \sqrt{{\hbar \omega_{k}}/({L c})}[\hat{a}_{k}(t)+\hat{a}^{\dag}_{k}(t)] $, $ L $ is the length of the cavity and $ c $ is its capacitance per unit length. Here $ \hat{a}_{k}(t) $ and $ \hat{a}^{\dag}_{k}(t) $ are the bosonic annihilation and creation operators for the $ k\textrm{th} $ mode, respectively. These operators satisfy the bosonic commutation relation $ [\hat{a}_{k}, \hat{a}^{\dag}_{j}]= \delta_{kj}$ and their time derivatives, in the Heisenberg picture employed here, are given by $ \partial_{t} \hat{a}_{k}=-i\omega_{k} \hat{a}_{k}$ and  $ \partial_{t}\hat{a}^{\dag}_{k}= i\omega_{k} \hat{a}^{\dag}_{k}$, where $\omega_{k}={k\pi}/({L \sqrt{\ell c}}) $ is the frequency of the $ k\textrm{th} $ mode in the cavity with $ \ell $ being the inductance per unit length. The current operator can be derived similarly and it is found to be $ \hat{I}_{\textrm{cav}} =\sum_{k=1}^{\infty} \hat{B}_{k}(t) \sin \left({k\pi x}/{L}\right) \eqqcolon\sum_{k=1}^{\infty}I_{k} $, with $ \hat{B}_{k}(t)  = {i}\sqrt{{L \hbar c\omega_{k}^{3}}/({ k^2 \pi^{2}})}[\hat{a}_{k}(t)-\hat{a}^{\dag}_{k}(t)] $.

The Hamiltonian of the cavity--resistor system in the absence of coupling is given by integrating the energy density in the cavity from $ 0 $ to $ L $ as $\hat{H}_{0} = \hat{H}_{\textrm{cav}}+\hat{H}_{R}=\int^{L}_{0} \! ({c\hat{V}_{\textrm{cav}}^{2}}+{\ell \hat{I}_{\textrm{cav}}^{2}})/2  \, dx+\hat{H}_{R}  $. We treat the coupling of the resistors to the cavity by adding their voltage fluctuations to the cavity voltage operator in $\hat{H}_{\textrm{cav}}$. Since the resistors are in practice orders of magnitude shorter than the photon wavelength they can be approximated as having a negligible spatial distribution and hence the potential drop across the ends of the resistor may be treated as a time--dependent discontinuity in the total voltage of the cavity at the position of the resistor. If the resistor is placed at $ x' $, the Hamiltonian for a cavity coupled to an embedded resistor is
\ba
&&\hat{H}(t) = \hat{H}_{R} + \frac{c}{2}\int^{x'}_{0} \! ( \hat{V}_{\textrm{cav}}\otimes \unity_{\textrm{res}}+\unity_{\textrm{cav}} \otimes \delta \hat{V}_{L})^{2} \, dx \nonumber \\  
&+& \frac{c}{2}\int^{L}_{x'} \! (\hat{V}_{\textrm{cav}}\otimes \unity_{\textrm{res}}+\unity_{\textrm{cav}} \otimes \delta \hat{V}_{R})^{2} \, dx + \frac{\ell}{2}\int^{L}_{0} \! \hat{I}_{\textrm{cav}} \, dx.
\ea
Here $ \delta \hat{V}_{L(R)} $ is the resistor-induced shift in the voltage on the left (right) side of the resistor and $ \unity $ is the identity operator. The tensor product, $ \otimes $, has been included to emphasise that $ \hat{V}_{\textrm{cav}} $ operates on the cavity and $ \delta \hat{V}_{L,R} $ on the resistor degrees of freedom. We neglect the fluctuations in the current due to the following reasons: The coupling between a resistor and the cavity modes decreases towards the ends of the cavity. Thus to achieve weak coupling the resistors are placed very close to the cavity edges.  Since the current operator vanishes at the edges and the energy contribution from the current is proportional to square of the current operator, this contribution is much smaller than the one arising from the voltage. Hence, the energy fluctuations are almost entirely detemined by the fluctuations in the voltage.

The Hamiltonian describing the interaction between the cavity and the resistor, $\hat{H}_\textrm{int}$, is found by calculating the difference in the energy of the cavity with and without the resistor. Neglecting second order terms in $\delta \hat{V}_{L,R}$ yields the analytic result
\be  \label{eq:Hint}
\hat{H}_\textrm{int} = \hat{H}(t)-\hat{H}_{0}= -\sum_{k=1}^{\infty} \frac{1}{\omega_{k}^{2}}\partial t \hat{I}_{k} \otimes \delta \hat{V}_{\textrm{res}} \eqqcolon  \hat{Q} \otimes \delta \hat{V}_{\textrm{res}},
\ee
where $ \delta \hat{V}_{\textrm{res}}=\delta \hat{V}_{L}-\delta \hat{V}_{R} $ is the total voltage fluctuation across the resistor and $ \hat{Q}(x')=-\sum_{k=1}^{\infty} {\omega_{k}^{-2}}\partial t \hat{I}_{k}(x') $ is the integral of the charge density in the cavity from $ 0 $ to $ x' $.
Note that $ \hat{H}_\textrm{int}(t) $ is of the form $\hat{Q} \otimes \delta  \hat{V}$,  where $ \hat{Q} $ acts on the system and $ \delta  \hat{V} $ on the Hilbert space of the environment. If the coupling strength is weak enough such that first order perturbation theory is valid and the autocorrelation time of the environment is sufficiently short as in a typical case, the transition rate between the cavity states $\ket{m}$ and $\ket{l}$ may be obtained from the Fermi golden rule as $\Gamma_{m\rightarrow l} \approx {|\bra{l} \hat{Q}\ket{m}|^{2}}{\hbar^{-2}} S_{V}(-\omega_{ml})$~\cite{Clerk10}. Here, $\omega_{ml}=({E_{l}-E_{m}})/{\hbar}$ corresponds to the energy change of the transition and $S_V (\omega)$ is the spectral density of environmental fluctuations causing the transition. The voltage fluctuations can be described as a semiclassical voltage source with the Johnson--Nyquist spectral power density $ S_{\delta V_{\textrm{res}}}(\omega) = {2 R \hbar \omega}/[{1-\exp(\frac{-\hbar \omega}{k_{B}T})}] $. In the case of two resistors, with the first resistor at $ x_{1} $ and the second at $ x_{2} $, the transition rates from the individual resistors can be added seperately since the coupling of the resistors to the cavity is linear and the voltage fluctuations of the resistors are not intrinsically correlated. Thus the transition rates of the different photon states in the cavity assume the forms
\be \label{eq:TUp}
\Gamma^{k}_{n\rightarrow n+1} = \frac{(n+1)\gamma^{(1)}}{\exp\left(\frac{\hbar \omega_{k}}{k_{B}T_{1}}\right)-1}+\frac{(n+1)\gamma^{(2)}}{\exp\left(\frac{\hbar \omega_{k}}{k_{B}T_{2}}\right)-1},
\ee
and
\be \label{eq:TDown}
\Gamma^{k}_{n\rightarrow n-1}= \frac{n\gamma^{(1)}}{1-\exp\left(\frac{-\hbar \omega_{k}}{k_{B}T_{1}}\right)}+\frac{n\gamma^{(2)}}{1-\exp\left(\frac{-\hbar \omega_{k}}{k_{B}T_{2}}\right)},
\ee
where we have defined $\gamma^{(i)}={2\omega_{1}}{R^{(i)}_\textrm{eff}}/({\pi Z_{0}})={2R^{(i)}_\textrm{eff}}/({L \ell})$, and an effective resistance $ R^{(i)}_{\textrm{eff}}=R_{i}\sin^{2}(\pi x_{i} /L) $. Here the characteristic impedance of the transmission line is $ Z_{0}=\sqrt{\ell/c}. $
 Intrinsic dissipation mechanisms can also be included by giving the cavity a small effective resistance per unit length $r$. This introduces a similar but far smaller term into the transition rate corresponding to the finite $ Q $ value of the bare cavity. Note that $ \gamma $ is inversely proportional to the total inductance of the cavity. Replacement of the center conductor with a series of superconducting quantum interference devices (SQUIDs) enables the tuning of $\ell $ independent of the other parameters. This would provide in situ control of the coupling strength and the transmission rates.


We aim to find the quasiequilibrium electron temperature of the second resistor, $T_{2}$,  and the transmitted photonic power in the case for which the temperature of the first resistor, $T_{1}$, as well as the phonon bath temperature $T$, are fixed. Experimentally, it is possible to use superconductor--insulator--normal-metal (SIN) tunnel junctions to both control and measure the  temperatures of the resistors \cite{Giazotto06,Timofeev09}.

The master equation for the occupation probabilities  $\vec{p}_{m}(t)=\{p^{m}_i(t)\}$ of a cavity mode $m$ can be expressed as $\frac{d\vec{p}_{m}(t)}{dt}= \Gamma^m \vec{p}_{m}(t)$. The transition matrix has elements $ \Gamma_{ik}^m= \Gamma_{k\rightarrow i}^m \text{ for }i \ne k$ and diagonal elements $\Gamma_{kk}^m= -\Gamma_{k \rightarrow k+1}^m - \Gamma_{k \rightarrow k-1}^m$. The stationary state of this system corresponds to the zero eigenvalue of matrix $\Gamma$ and yields the probability distribution in the quasiequilibrium.

To find the temperature of the second resistor we need to know only the net photon power transferred from the cavity to the second resistor, $P^{(2)}_{\textrm{cav}}$, and the power leaving the electron cloud of the resistor to the phonon bath due to electron--phonon coupling, $P^{(2)}_{\textrm{el-ph}}$. Here, we assume that other heat conduction mechanisms, such as the quasiparticle heat conduction, are negligible. In equilibrium, there is no net heat transfer from the resistor and hence, $ P^{(2)}_{\textrm{el--ph}}=P^{(2)}_{\textrm{cav}}  $. The electron--phonon coupling is given by $ P^{(2)}_{\textrm{el-ph}} = \Sigma V (T_{2}^{5}-T^{5}) $ 
where $\Sigma$ is a material--dependent constant and $V$ is the volume of the resistor \cite{Giazotto06}.
Thus the quasiequilibrium temperature  of the second resistor is given by
\be \label{eq:T2eq}
T_{2}=\sqrt[5]{{P^{(2)}_{\textrm{cav}}}/({\Sigma V})+T^{5}}.
\ee
Since $P^{(2)}_{\textrm{cav}}$ is the rate at which energy leaves the cavity due to absorption of photons by the second resistor, this is just equal to the energy of a single photon multiplied by the rate at which the number of photons in the cavity decreases due to Resistor 2. For each mode, this quantity is the  probability of having $ n $ photons in the mode
multiplied by the energy difference in the transition and the net transition rate for that mode. To obtain the total power, we sum over all the modes as
\be \label{eq:pcav}
P^{\textrm{net}}_{\textrm{cav}} = \sum_{k}\hbar \omega_{k} \sum_{n}\left(\Gamma^{(2),k}_{n \rightarrow n-1}- \Gamma^{(2),k}_{n \rightarrow n+1}\right)p^{k}_{n}.
\ee
In this paper $\Gamma^{(m),k}_{n \rightarrow n\pm 1}$ represents the transition rate for the $ k \textrm{th}$ mode due to the $ m \textrm{th}$ resistor. These are given by Eqs.~(\ref{eq:TUp}) and (\ref{eq:TDown}) by keeping only the term corresponding to Resistor $ m $. The first sum is over the modes and the second is over the photon number. Since the transition rates depend on $ T_{2} $, Eqs.~(\ref{eq:T2eq}) and (\ref{eq:pcav}) must be solved self-consistently. The effective temperature of the cavity, $T_\textrm{eff}$, can be defined in the spirit of the detailed balance as  \cite{Clerk10} \be  \label{eq:tef1}
T_{\rm{eff}}=\frac{\hbar \omega_{1}}{k_{B}\log \left( \frac{\Gamma_{0\rightarrow 1}}{\Gamma_{1\rightarrow 0}}\right)}.
\ee


The parameters used are experimentally feasible: The effective resistance and capacitance per unit length of the cavity are $r=2 \times 10^{-3}  \textrm{ }\Omega \textrm{m}^{-1} $ and $c= 130\times 10^{-12}\textrm{ Fm}^{-1}$, respectively. The length of the cavity is $ L=6.4~\textrm{ mm} $ with $ \omega_{k}=2\pi k \times 10^{10} \textrm{ s}^{-1}$.  These values are consistent with the resonators fabricated in Ref.~\cite{Wallraff08} and give a charcteristic impedance of $Z_{0}= 60.1\textrm{ }\Omega $.  Resistors 1 and 2 are identical with volume $V=2.25 \times 10^{-20} \textrm{ m}^{3}$ and are placed symmetrically in the cavity, offset from the edge by $ 0.1\times L $ and $  0.9\times L$, respectively. We employ the parameters of Au$_{0.75}$Pd$_{0.25}$ with the electron--phonon coupling constant $\Sigma= 3\times10^{9} \textrm{ Wm}^{-3} \textrm{K}^{-5}$. Similar resistors were produced in Ref.~\cite{Timofeev09}. We compute the equilibrium temperature for resistances of $  R_{1}=R_{2}=R=230\textrm{ } \Omega $ and $ R=2.3 \textrm{ } \Omega $ yielding $ \gamma = 1.53 \times 10^{9}\textrm{ s}^{-1}$ and  $1.53 \times 10^{7}\textrm{ s}^{-1} $ respectively.

\begin{figure}[hbtp!]
	\includegraphics[width=0.4\textwidth]{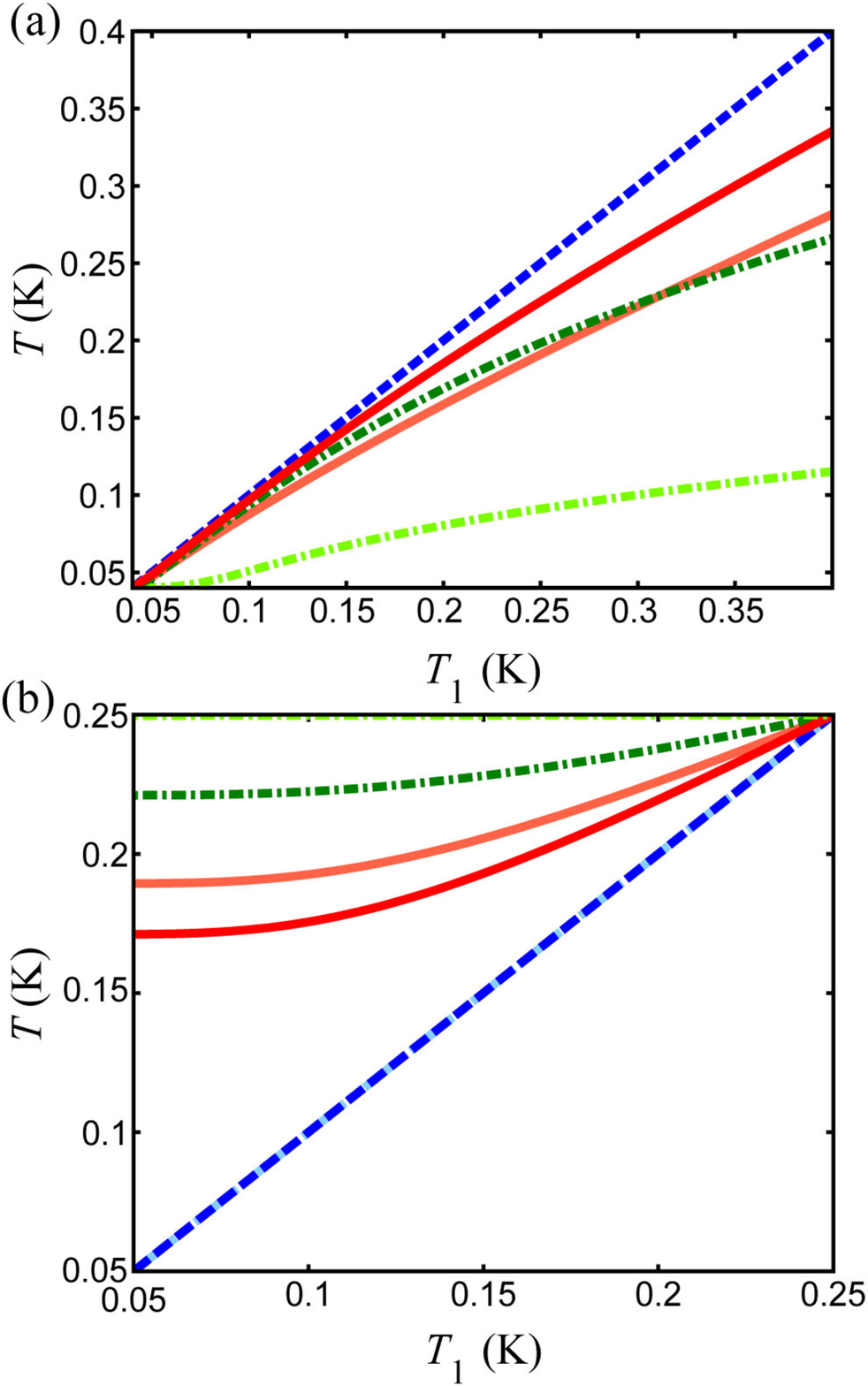}

	\caption{(color online) Temperature of Resistor 2 (dash-dotted line) and the effective temperature of the cavity (solid line) plotted against the temperature of Resistor 1. The temperature of Resistor 1 (dashed line) is also shown for reference. The darker curves were obtained for $R=230\textrm{ } \Omega $ and the lighter curves for $R=2.3\textrm{ } \Omega $. In panel (a), single-photon heating is observed for a phonon bath temperature, $ T=40\textrm{ mK}$. In panel (b), single-photon cooling is observed for $ T=250\textrm{ mK}$. }
\label{fig:mainresult} 
\end{figure}

The numerical results are shown in Fig.~\ref{fig:mainresult}(a). At $ 40 \textrm{ mK} $ bath temperature, $ T_{1} $ is changed from $ 40  \textrm{ mK} $ to $ 400  \textrm{ mK} $. With $R=230\textrm{ } \Omega $ and at low $ T_{1} $, the temperature of Resistor 2 closely follows that of Resistor 1 since the single-photon channel dominates over all other heat conduction mechanisms. The heat flow out from the hot electrons in Resistor 2 due to the electron--phonon coupling increases rapidly with the temperature, and hence we observe a clear deviation of $T_2$ from $T_1$ above $250~\textrm{mK}  $. With $ R=2.3\textrm{ }\Omega$ this heating effect is still clearly observable although weaker due to the much weaker coupling.

With a bath temperature of $ 250 \textrm{ mK} $, single-photon cooling of Resistor 2 is shown in Fig.~\ref{fig:mainresult}(b). With $R=230\textrm{ } \Omega  $ the temperature of Resistor 1 is lowered from $250~\textrm{ mK}  $ to $50~\textrm{ mK}  $ and $ T_2 $ saturates to roughly $225~\textrm{ mK}  $. At such bath temperatures the electron--phonon coupling makes a significant contribution to the power transfer between the resistors and the bath. On the other hand, as the temperature of Resistor 1 decreases it excites exponentially fewer photons in the cavity and thus the photonic power exhange between the resistors saturates. As a consequence of the increased bath temperature, single-photon cooling is particularly susceptible to any reduction in the coupling, and hence is very low for $ R=2.3\textrm{ }\Omega$.

By altering either the resistance or the offset of the resistors from the edge of the cavity the coupling can be tuned. The stronger the coupling, the greater the change in the temperature of the Resistor 2, i.e., $ T_{2} $ follows $ T_{1} $ more closely at higher temperatures. However, our assumption of weak coupling breaks down beyond a certain limit, namely, the effective $Q$ value of the cavity must be much greater than unity. 
For $R=230\textrm{ }\Omega$, we have $Q^{\,\textrm{eff}}\approx\omega_{1}/(2 \gamma)=20.5$.

We also analyzed the system semiclassically, employing the repeated lumped element model shown in Fig.~\ref{fig:CPW}(b) but without using quantum operators.
The transmission line is divided into $ N $ points each with a capacitor of size $ c\Delta x $ and an inductor of size $ \ell \Delta x$. We take $ N=100$ which is sufficiently large for the discrete model to give a good approximation of the continuous transmission line.  At two of these points $ k= \{i,j\}$, the inductors are replaced with resistors and the corresponding voltage fluctuations are introduced. Applying Kirchoff's current and voltage laws to the $ n\textrm{th} $ block in the circuit and eliminating the voltage gives, in Fourier space, the system of equations $\emph{Z}(\omega)\vec{I}(\omega)=\delta \vec{V}(\omega) $, where $ \emph{Z}(\omega) $ is a tridiagonal matrix which accounts for the impedance in the transmission line, $ \vec{I}(\omega) $ and $\delta \vec{V}(\omega) $ are vectors containing the current and voltage fluctuation at each point of the line.
We hold Resistors 1 and 2 at constant temperatures and assume independent noise fluctuations. This allows us to calculate the change in the current due to the noise from each resistor seperately. The time-averaged noise power
from Resistor $ i $ to Resistor $ j $ can be found directly from the current as
\be \label{eq:ClassPow}
P_{i \rightarrow j} = \frac{R_{j}R_{i}}{\pi} \int_{-\infty}^{\infty}  \!  \left(\frac{\hbar \omega |(Z^{-1})_{ji}|^{2}}{1-\exp(\frac{-\hbar \omega}{k_{B}T_{i}})}-\frac{\hbar \omega |(Z^{-1})_{ij}|^{2}}{1-\exp(\frac{-\hbar \omega}{k_{B}T_{j}})}\right) \, d\omega,
\ee
which can be calculated numerically for a given $ T_{i} $ and $ T_{j} $. It should be noted that Eq.~(\ref{eq:ClassPow}) has the same form as the photonic power given in Ref.~\cite{Cleland04}.

\begin{figure}[hbtp!]
{
\includegraphics[width=0.4\textwidth]{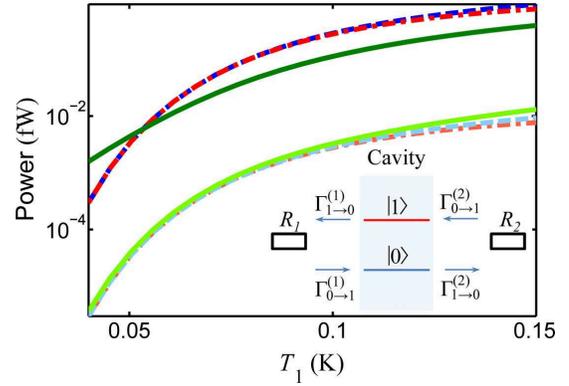}

}
\caption{(color online) Quantum (broken lines) and semiclassical power (solid line) from Resistor 1 to Resistor 2 for $ R=2.3 \textrm{ }\Omega $ (bottom lines) and $ R=230 \textrm{ }\Omega $ (top lines). The temperature of the first resistor spans the range from $40\textrm{ mK}$ to $150\textrm{ mK}$ while the second resistor is held at at $T_{2}= T_{1}-20 \textrm{ mK} $. The full numerical simulation (dashed line) which includes all the modes is in good agreement with the two-level approximation (dash-dotted line) in this temperature regime. Inset: illustration of the two-level model for the system of Fig.~\ref{fig:CPW}(a).}
\label{fig:CompPow}
\end{figure}

In the quantum case with $ k_{B}T \ll \hbar \omega_{1}$, it is reasonable to approximate the cavity as a two-level system composing of zero or one photons in the fundamental mode. In this limit the photonic power transfer can be solved analytically if both $T_{1}$ and $T_{2}$ are known. The inset of Fig.~\ref{fig:CompPow} illustrates the process schematically. Since the probabilities must sum to unity, $p_{0}+p_{1}=1$, we have $ p_{0}=\frac{\Gamma^{-}}{\Gamma^{\Sigma}}$ and  $  p_{1}=\frac{\Gamma^{+}}{\Gamma^{\Sigma}}$, where $\Gamma^{-}= \Gamma_{1 \rightarrow 0}^{(1)}+\Gamma_{1 \rightarrow 0}^{(2)}$, $\Gamma^{+}= \Gamma_{0 \rightarrow 1}^{(1)}+\Gamma_{0 \rightarrow 1}^{(2)}$, and $\Gamma^{\Sigma}= \Gamma^{-}+\Gamma^{+}$. Therefore the net quantum photon power transferred out of the second resistor is
\be \label{eq:QuantPow}
P_{Q}=\Gamma_{2}^{+}p_{0}\hbar \omega_{1} - \Gamma_{2}^{-} p_{1}\hbar \omega_{1} = \frac{\hbar \omega_{1}}{\Gamma^{\Sigma}} \left(\Gamma_{2}^{+}\Gamma^{-}- \Gamma_{2}^{-} \Gamma^{+}\right),
\ee
which can be expressed in terms of the physical quantities using Eqs.~(\ref{eq:TUp}) and (\ref{eq:TDown}).

The results of Eqs.~(\ref{eq:ClassPow}) and (\ref{eq:QuantPow}) are plotted in Fig.~\ref{fig:CompPow} alongside the full numerical treatment of the quantum power. The temperatures of both resistors are fixed here, with the temperature of Resistor 2 held $20\textrm{ mK}$ below the temperature of Resistor 1 which is increased from $40\textrm{ mK}$ to $150 \textrm{ mK}$. With $R=2.3\textrm{ }\Omega $ there is very close agreement between quantum and semiclassical power even as the power output changes over several orders of magnitude. At high resistances comparable to $ 230\textrm{ }\Omega $, the quantum and semiclassical models have a somewhat different temperature dependence.
 \FloatBarrier
In summary, we have presented analytical and numerical results which show that heat conduction by single photons has a distinct signature which is readily observable in the heat transport between two resistors coupled by a superconducting cavity.  Photonic heat conduction enables the possibility to perform remote heating or cooling of low-temperature circuit elements at an extremely narrow bandwidth, thus having a minimal perturbation to the usual operation of the component. Furthermore, we showed that although weak, the coupling of resistors to the superconducting cavity can dominate the dissipation mechanisms, and hence completely determine the temperature of the cavity, i.e., these resistors work as engineered and controllable environments for the cavities. Thus our work shows that SIN thermometry and temperature control can be extended to superconducting cavities. We observed a disagreement between the quantum and semiclassical models for relatively strong coupling. This calls for further theoretical and experimental studies on the true quantum nature of single--photon heat conduction. We thank Jukka Pekola and Teemu Ojanen for helpful discussions, and Academy of Finland and Emil Aaltonen Foundation for financial support.

\bibliographystyle{apsrev}
\bibliography{RefList}
\end{document}